\documentclass[12pt]{article}
\usepackage{amsmath,amssymb}

\def\bea{\begin{eqnarray}}
\def\ben{\begin{equation}}
\def \een{\end{equation}}
\def \eea{\end{eqnarray}}
\def \bp{{\bf}}\def \cE{{\cal E}}

\def \p{\partial}

\def\beq{\begin{eqnarray}}
\def\eeq{\end{eqnarray}}

\def\d{\partial}

\def \bp {{\bf p}}

\newcommand{\bg}{\begin{gather}}

\newcommand{\bseq}{\begin{subequations}}
\newcommand{\eseq}{\end{subequations}}

\renewcommand{\tanh}{\mathop{\rm th}\nolimits}

\renewcommand{\ln}{\mathop{\rm ln}\nolimits}

\def\half{\frac{1}{2}}

\begin{document}

\title{\textbf{A Spacetime Geometry picture  of Forest  Fire Spreading
and of Quantum Navigation}}

\vspace{2cm}
\author{\bf G.W. Gibbons}

\date{}
\maketitle
\begin{center}
\hspace{-0mm}
{\bf  D.A.M.T.P.}
\hspace{-0mm}
\end{center}
\begin{center}
\hspace{-0mm}
 {\bf University of Cambridge}
\hspace{-0mm}
\end{center}
\begin{center}
\hspace{-0mm}
{\bf Wilberforce Road, Cambridge  CB3 0WA }
\hspace{-0mm}
\end{center}
\begin{center}
\hspace{-0mm}
{\bf U.K }
\hspace{-0mm}
\end{center}
\vspace{2cm}

\begin{abstract}
  The problem of finding null geodesics in a stationary Lorentzian spacetime
  is known to to be equivalent to finding the geodsics of a  Randers-Finlser
  structure. This latter problem is equivalent to finding the motion of charged particles
  moving on a Riemannian manifold in a background magnetic field or equivalently, by a generalization
  of Fermat's principle,  to  Zermelo's problem of  extremizing travel  time
  of an aeroplane in the presence of a wind. In this paper this triad of equivalences
  is extended to include  recent model of the spread of a forest fire  which uses  form of Huyghen's principle.
  The construction may also be used to solve  a problem in quantum control theory in which one seeks a control Hamiltonian
  taking an initial state of a quantum mechanical system with its own Hamiltonian to a desired final state in least time.
  The associated stationary spacetime may be thought of as defined on an  extended quantum phase space
  (Souriau's evolution space), the space of quantum stares being complex projective space
  equipped with its Fubini-Study K\"ahler metric. It is possible that this spacetime view point
  may provide insights relevant  for our understanding of  quantum gravity.

\end{abstract}

\vskip 2 cm
\noindent
\rule{7.7 cm}{.5 pt}\\
\noindent 
\noindent
\noindent ~~~ {gwg1@damtp.cam.ac.uk}

\newpage
    \tableofcontents
\pagebreak

\newpage

\section{Introduction}
Over the last few years it has emerged
\cite{Robles,Gibbons:2008zi,Javaloyes:2013ika,Caponio:2014gra}  
that there is a close
connection between three superficially unrelated problems.
\begin{itemize}

\item The motion of light rays in a stationary spacetime.

\item A particular kind of Finsler geometry  due to   Randers  \cite{Randers:1941gge}. 

\item A variational problem generalising  the notion of
  a geodesic of a Riemannian  metric due to Zermelo \cite{Zermelo}.  

\end{itemize}

The interest of these connections goes beyond General Relativity: there
are applications to a number of other  areas of physics.
For instance to the propagation of sound in a wind \cite{Gibbons:2009ts,
Gibbons:2011ib}, the behaviour of quasi-particles in graphene sheets \cite{Cvetic:2012vg}, problems in quantum  control theory 
\cite{R1,R2,Brody:2014voa,BM2,RS3,Brody:2014jaa,Brody:2015zra} and most recently the spread of wild fires \cite{Markvorsen}. 
This  latest, rather surprising,   application makes use of Huyghen's principle and this motivates   
the present paper the aim of which is to both explore in more detail this aspect from  
from a spacetime point of view
but also to expand upon from the same standpoint the results of \cite{Brody:2014jaa,Brody:2015zra}.
The  common idea   will be the systematic
exploitation of  the Hamilton-Jacobi and Eikonal equations.

\section{Randers-Finsler metrics, Zermelo's problem and null geodsics}  

In the interest of making this paper self-contained
we start with the a brief review of the three legs of the correspondence
described in \cite{Gibbons:2008zi}. It will also serve to establish our notation and conventions.

\subsection{Finsler geometry} 
In Riemannian  geometry the function on the tangent bundle given by   
\ben
F= \sqrt{ a_{ij}(x)  v^i v^j } \label{Finsler}  
\een
where $a_{ij}$ is the Riemannian metric,
is homogenous degree one in the fibre coordinate $v^i$ and serves
to define a norm \footnote{sometimes, confusingly for a physicist,
called a Minkowski norm} on the tangent space at each point.
Moreover one may define geodesics as critical points
of the length  functional on curves $x^a(\lambda)$ with tangent vectors
$v^i = \frac{dx^i}{d \lambda}$ given by 
\ben
S[x^i (\lambda)] = \int F d \lambda \,.
\een
Note that the functional $S[x^i(\lambda)]$ is invariant
under reparametrisations $\lambda \rightarrow g(\lambda)$.
In Finsler geometry one replaces  the right hand side of (\ref{Finsler})
by any any homogeneous function of degree one of the tangent vector
provided it defines a norm.
The co-dimensions one convex surface in the tangent space given
by
\ben
F =  1 \,,
\een
is called the indicatrix. For the Riemannian case it is an ellipsoid
and is centro-symmetric. In general the indicatrix
of a Finlser structure is neither an ellipsoid nor centro-symmetric.

The special case we shall be most interested in is
that of
Randers for which
\ben
F= \alpha + \beta \,, \qquad \alpha =\sqrt{ a_{ij}v^iv^j } \,,\qquad \beta= 
b_i v^i \,.\een
The indicatrix is an ellipsoid but displaced from the origin
by $b_i$. Thus we require that its length with respect to the
Riemannian   metric $a_{ij}$, i.e. $\sqrt{a^{ij}a_ia_j}$  be less than unity.
Randers-Finsler geodesics  satisfy
\ben
\frac{d^2 x^i}{ds ^2} + \Big \{{\, _j } {\,{^i} \,}{\, _k} \Big \}
\frac{dx^j}{ds} \frac{dx^k}{ds} = a^{ij}F_{jk}  \frac{dx^k}{ds}\,,  
\een
where $ds=\alpha d \lambda $ is arc length with respect to the
Riemannian metric $a_{ij}$,    $ \Big \{{\, _j } {\,{^i } \,}{\, _k} \Big \}$
are the Christoffel symbols of the Riemannian metric $a_{ab}$ 
and $F_{ij} = \p_i b_j-\p_j b_i $. 
If $F_{ij}$ vanishes then  Randers-Finsler geodesics are just
ordinary geodesics of the Riemannian  metric $a_{ij}$. If $F_{ij}\ne0 $
then they are  what are sometimes called magnetic geodesics, i.e.  the paths
of charged  particles moving in a magnetic field $F_{ij}$.
Note that while two Randers-Finsler  structures 
with one-forms  $b_i$ and $\tilde b_i$ for which
\ben
b_i-\tilde b _i = \p_i h\,,
\een
for  some function $h$, have the same  Randers-Finsler geodsics
they are nevertheless  not equivalent as Randers-Finsler structures.

\subsubsection{The Funk-Randers  metric}
The fore-going remarks may be illustrated
by the following celebrated example

Given a convex domain in Euclidean space Funk introduced
a distance function $d(x,y)$  which is asymmetric $d(x,y) \ne d(y,x)$
which generalises a  standard construction of a symmetric distance
function by Cayley in the case of the unit ball. In general the symmetrisation
$\half \bigl ( d(x,y)+d(y,x)  \bigr )$ of the Fink distance  is called  
the Hilbert distance.  In what follows
we shall specialise to  the case of the unit ball.
In that case the Funk  distance function arises from a  Randers Finsler metric
for which \cite{Okada,Kozma}
\ben
F=
\frac{\sqrt {(1-x^ix^i)  v^j v^j +(v^j x^j)^2 }+ (x^k v^k) }
{1- x^l x^l}        \,.  
\een
Evidently since
\ben
\p_i b_j-\p_j b_i =0\,,
\een
the Finsler geodsics are unaffected by
the one-form $b_a$ so the Finsler geodesics coincide with the geodesics of the
symmetrized   Finsler  function, the metric of which is given by 
\ben
a_{ij} dx^i dx^j = \frac{dx^i dx^i (1 -x^k x^k)  +( dx^jx^j )^2 }
{ \bigl (1- x^lx^l  \bigr ) ^2}\,.   
\een
In  polar coordinates $x^i=r(\cos \phi,\sin \phi)$  
\bea
ds ^2 &=& \frac{dr^2}{(1-r^2)^2 } +  \frac{r^2}{1-r^2} d \phi ^2\\
 &=& \frac{1}{4} \Bigl \{  d \psi ^2 + \sinh^2 \psi d \phi ^2  \Bigr \} \,,   \eea
where $r=\tanh \frac{\psi}{2}$. 
This is a metric of constant curvature $-\frac{1}{4}$ and the coordinates
$x^a$ are Beltrami coordinates in which the geodesics are straight lines
(see e.g. \cite{Gibbons:2006mi}).
One therefore has
\ben
b_idx^i = \half d \ln \cosh \frac{\chi}{2} = d \frac{1}{\sqrt{1-r^2}}\,. 
\een

The full Funk distance function may be expressed as
\ben
d(x,y)= \half d_{\mathbb{H}^2 } +
\frac{1}{\sqrt{1- x^ix^i }} - \frac{1}{\sqrt{1- y^j y ^j }} \,,  
\een
where  $d_{\mathbb{H}^2 }$ is the standard Hyperbolic distance
function on the unit disc and may be expressed in Beltrami-coordinates as  .
\ben
d_{\mathbb{H}^2 } = \cosh ^{-1} \Big (\frac{1-x^iy^i  }
{\sqrt{ 1-x^jx^j} \sqrt{1-y^ky^k} }       \Bigr )  \,.
\een

\subsection{The Zermelo Problem}
An example of  Zermelo's problem is faced by  
the  pilot of an  aircraft flying in the
presence of a wind  of speed $W^i$ relative to the ground,
seeks to minimize journey time, given that the aircraft
flies at constant speed relative to the air. If the journey is of
any distance, for example a transatlantic flight, the pilot must
take into account the curvature of the earth and so
in the absence of  wind the optical route would be a great circle,
i.e a geodesic of a Riemannian manifold with metric $h_{ij}$ .
The pair $\{ h_{ij}, W^i\}$ are referred to as the Zermelo data.

According to \cite{Robles} this problem is  equivalent to finding the
shortest length 
of a curve in a manifold equipped with a  Randers-Finsler metric
with Randers data $\{a_{ij}, b_i\}$. The equivalence
and the relationship between the two sets of data is conveniently
expressed in terms of a third problem, finding the motion
of a light ray moving in a time-independent Lorentzian spacetime.
\subsection{Stationary Metrics}
A  general stationary Lorentzian  metric may be expressed as  
\ben
g_{\mu \nu} dx^\mu dx ^\nu  = -
V^2 \bigl(dt + \omega _idx^i \bigl ) ^2  + \gamma _{ij} dx ^i dx ^j
\label{spacetime}\,,\een
where   $V, \omega _i, \gamma_{ij} $ are independent  of the time
coordinate $t$. In general the time coordinate $t$ is not  unique
since one  may always make  the replacement 
\ben
t \rightarrow  t -f(x^i) \,,
\qquad \omega_i \rightarrow \omega _i + \frac{\p f}{\p x^i} \,..   
\een
Locally at least one may regard the spacetime as an $\mathbb{R}$ bundle over
the spatial manifold, i.e. the space of orbits of the
Killing vector field $\frac{\p}{\p t}$,
and $\omega_i$ are the components of the horizontal connection,
known in this context
as the Sagnac connection \cite{Ashtekar:1975wt}.
If the curvature $F_{ij}=\p_i  \omega_j= \p_i \omega_j $ vanishes
one may, at least locally, choose $f$ to make $\omega_i$ vanish.
Such spacetimes are said to be static and admit an extra time-reversal  symmetry
$t \rightarrow -t$.

Evidently we have 
\ben
g_{ij}= \gamma _{ij} -V^2 \omega _i \omega _j\,.
\een

Fermat's princple arises from  the Randers structure given by
\bea
a_{ij}&=& V^{-2} \gamma _{ij}\,,\\ 
b_i & =& -\omega _i \,.
\eea

This is equivalent to a Zermelo structure of the form
\bea
h_{ij}&=& {1 \over 1 + V^2 g^{lm} \omega _l \omega _m } { g_{ij} \over
  V^2 }\,, \\
W^i&=& V^2 g^{ij}\omega _j \,. 
\eea

We have
\bea
\gamma _{ij}&= & g_{ij} +V^2 \omega _i \omega _j\,,\\
\gamma ^{ij} &=& g^{ij} - { V^2 \omega ^i\omega ^j\over 1+ V^2
  g^{lm} \omega_l \omega _sm} \,, 
\eea
with
\ben
\omega ^i = g^{ij} \omega _j \,,\qquad W^i ={  V^2 \gamma ^{ij} \omega
  _j
\over 1-V^2 \gamma ^{lm} \omega _l \omega _m } \,,
\een
\ben
1- V^2 \gamma ^{ij} \omega _i \omega _j= 
{1 \over 1+ V^2 g^{ij} \omega _i \omega _j}  \,.
\een

Note that
\ben
|W^i|^2 = h_{ij}W^i W^j = a_{ij} b^i b^j  = | b|^2 \,.
\een

In terms of the the Zermelo data the spacetime metric (\ref{spacetime})   is  
\ben
ds ^2 = \frac{V^2}{1-h_{ij} W^i W^j }\bigl[ - dt ^2 +
  h_{ij}(dx^i-W^i dt) ( dx^j-W^j dt )
 \bigr ] \,.\label{Zspacetime}
\een

Note that both the Randers-Finsler structure  and the Zermelo
structures do not depend upon the conformal equivalence class
of the metric (\ref{spacetime}). In other words they are unchanged
by the replacements (known as  Weyl rescaling)  
\ben
V^2 \rightarrow \Omega ^2 V^2 \,, \qquad \gamma_{ij} \rightarrow \Omega ^2
\gamma_{ij}\,,\qquad \omega_i  \rightarrow \omega _i \,, \label{Weyl}
\een  
where $\Omega$ is an arbitrary non-vanishing function of both
the space \emph{and}  time coordinates. 

Note also that under (\ref{Weyl} ) we gave 
\ben
 g_{ij} \rightarrow \Omega ^2  g_{ij} \,,\qquad  
g^{ij} \rightarrow \Omega ^{-2}  g^{ij} \,.
\een
Clearly therefore, we may pick $\Omega$ to be given by
\ben
\Omega^2  = 1- h_{ij} W^i W^j \,,
\een 
in which case (\ref{spacetime}) or equivalently
(\ref{Zspacetime}) may be replaced by the metric 
\ben
 - dt ^2 + h_{ij}(dx^i-W^i dt) ( dx^j-W^j dt ) \,. \label{Lorentz}
\een

\subsection{Wave Fronts}

If one thinks of light propagation
and adopts a covariant spacetime perspective,
the associated  wave fronts are null hypersurfaces of spacetime 
$S(x^\mu) = {\rm constant}$ satisfy
\ben
g^{\mu \nu} \partial  _\mu S \partial _\nu S  =0\,,
\qquad \frac{ dx^\mu}{d \lambda} = g^{\mu \nu} \partial _\nu S \,. \label{null1}  
\een
and the associated light rays $x^\mu=x^\mu(\lambda)$  are
their null generators satisfying 
\ben
g^{\mu \nu} \p _\mu S \p_\nu =0\,,
\qquad \frac{ dx^\mu}{d \lambda} = g^{\mu \nu}\p_\nu S \,. \label{null2} 
\een
Four statements  follow directly from (\ref{null1}, \ref{null2}).
\begin{itemize}
\item The  rays lie in the wave fronts
\ben
\frac{dS}{d\lambda}= \frac{dx^\mu }{d \lambda} \p_\mu S =0 \,.
\een
\item  The the light rays have  null tangent vectors:
\ben
g_{\mu \nu} \frac{dx ^\mu }{d \lambda} \frac{dx ^\nu }{d \lambda} =0\,.
\een
\item
The light rays  are geodesic
\ben
\frac{dx ^\nu}{d \lambda} \nabla _\nu \frac{d x^{[\mu} }
{d \lambda } \frac{dx^{\sigma  ]} }{\d \lambda}  =0\,,
\een
where $\nabla_\mu$  is the covariant derivative with respect
to the Levi-Civita associate to the metric $g_{\mu \nu}$.  

\item
The parameter  $\lambda $ is an affine parameter
\ben
\frac{dx ^\nu}{d \lambda} \nabla _\nu \frac{d x ^\mu} {d \lambda }   =0\,. 
\een
\end{itemize} 
Only the last statement is not invariant under  conformal rescaling
of the spacetime metric
\ben
g_{\mu \nu} \rightarrow \Omega ^2 g_{\mu \nu} \,. 
\een

\subsubsection{Randers viewpoint}

In the Randers  framework equation (\ref{null1}) becomes   
\ben
a^{ij} \Bigl( \partial_i S + b_i \partial _r S\Bigr )
\Bigl ((\partial _j S + b_j \partial _t S \Bigr)  =  \Bigl( \partial_t   S\Bigr )^2   
\,.\een

Separating off the time-dependence 
\ben
S=-t + W (x^i) 
\een
yields
\bea
a^{ij}\bigl((\p _iW -b_i \bigr )\bigl( \p_i W -b_i \bigr )&=& 1\,,\\
\frac{dx^i}{{d \lambda}}&=& a^{ij} ( \p_jW -b_j)    \,,\\
\frac{dt}{d \lambda} &=&(1 -    a^{ij}b_ib_j ) + a^{ij} b_i \p_j W \,.   
\eea

\subsubsection{Zermelo viewpoint}

In the Zermelo framework the Hamilton-Jacobi equation becomes
\ben
h^{ij} \p_ iS \p_j S - (\p_tS + W^i\p_i S) ^2 =0\,,   
\een
which may be rewritten, taking the positive square root,  as
\ben
\p_t S = \sqrt{(h^{ij}\p_iS \p_i S )} -W^i \p_i S  
= G(x,p_i) = \sqrt{h^{ij}p_ip_j} -W^ip_i \,,\label{HJred}
\een
where $p_i=\p_iS$. Now  $G$  coincides with
the expression given in  equation (14) of \cite{Gibbons:2008zi}
for  the moment map  generating  the Zermelo flow on the cotangent space
of the spatial manifold.

In fact equation (\ref{HJred})  is in striking
agreement with the Hamilton-Jacobi equation obtained in section \S 2.2
of \cite{SGRMC} by entirely different arguments.

 Further progress typically depends upon being able to continue to 
separate variables
either because  further ignorable coordinates  arise from Killing vector 
fields  generating isometries of the metric $h_{ab}$ or 
because  of the existence of higher rank Killing tensor fields.
Note that by Froebenius's theorem, we may only introduce as many
ignorable coordinates as there are mutually commuting Killing vector fields.

\subsection{Winds which are Killing vector fields}

Consider the metric
\ben
-dt^2 + h_{ij}(dx^i-W^idt) (dx^j-W^jd t) = g_{\mu \nu} dx^\mu dx^nu
\een
Using Hamilton-Jacobi theory one has that null geodesics 
are given by
\ben
\frac{dx^\mu}{d\lambda} = g^{\mu \nu} \p _\nu S\,,\qquad 
g^{\mu \nu}\p_\mu S \p_\nu S =0\,,
\een
where $\lambda$ is an affine parameter
and $i=1,2,\dots m$.
Since
\bea
g_{\mu\nu} &=&  \begin{pmatrix} -1+ W^2 & -W_i \\ -W_i & h_{ij}  \end{pmatrix}\\
g^{\mu\nu} &=&  \begin{pmatrix} -1& -W_k \\ -W^j  & h^{jk} - W^jW ^k  \end{pmatrix} 
\eea
with $W^2=h_{ij} W^i W^j$, and $W_i=h_{ij}W^j$ etc.
In the present case the second equation become
\ben
(\p_t S + W^i \p_i S) ^2 = h^{ij} \p_i S \p_j S   
\een
If $S=-t + F(x^i) $ we have
\ben
(E  -  W^i \p_i F) ^2 = h^{ij} \p_i F \p_j  F   
\een

and 
\bea
\frac{dt }{d \lambda} &=&E - W^k \p_k F \\
\frac{dx^i} {d \lambda} &=& E W^i + (h^{ik} -W^i W^k) \p_k F \,.
\eea

Now suppose that $W^i$ is a Killing vector of $h_{ij}$. We may find adapted 
coordinates $\phi, x^A$, $A=2,3,\dots,m$ such that 
$W^i=\Omega \delta^i_\phi$   
where $\Omega $ is a constant. The metric $h_{ij}$ thus takes the
form
\ben
h_{ij}dx^idx^j= h_{AB} dx^A dx^B + 2 h_{A \phi} dx^A d \phi +
h_{\phi \phi}d \phi ^2 \,. 
\een
That is, $h_{ij}$  is independent of $\phi$ and $\phi$ is thus 
and ``ignorable'' or ``cyclic'' coordinate.

We may set $S=-Et + J \phi + W(x^A) $, with $J$ a constant 
and find 
\ben
(E  - \Omega J ) ^2 = h^{ij} \p_i F \p_ jF \,. 
\een

Thus
\ben
\frac{1}{|E-\Omega J |} F \,, 
\een
satisfies the Hamilton-Jacobi equation of the metric $h_{ij}$  
and hence 
\ben
T^i= \frac{1}{|E-\Omega J |} h^{ij} \p _j F  
\een
is a unit tangent vector to a geodesic.
We have 
\ben
\frac{dt}{d \lambda} = |E-\Omega J| \,,
\een
and 
\ben
\frac{dx^i} {d \lambda} =   ( E- \Omega J)  W^i  + h^{ik} \p_k F \,. 
\een
That is 
\ben
\frac{1}{|E-\Omega J|}  \frac{dx^i} {d \lambda} = W^i + T^i \,,
\een
whence 
\ben
 \frac{dx^i} {d t} = W^i + T^i \,,
\een
where $T^i$ is a unit tangent vector to a geodesic of the metric $h_{ij}$.

\section{Forest Fires} 
In the  remarkable recent paper \cite{Markvorsen}  
it has been pointed that the equations of  Richards
for large scale wildfire  spreads (see e.g. \cite{SGRMC})  may be cast in
terms  of a  Finsler function $F(x,y,u,v)$  giving the distance
between successive firelines as the fire advances
according to Huyghen's principle. Here $(x,y)=x^i$ are
spatial coordinates and  $(u,v)=v^i$
velocities with $i=1,2$ and $F(x^i,v^i)$ is homogeneous degree one in $v^i$
:$ F(x^i,\lambda v^i)  =\lambda F(x^i,v^i)$. A particular
example, known  as the
hemi-spherical elliptic model,  discussed  in  \cite{Markvorsen}  is  of
Randers type. 
It belongs to
a slightly  wider class of  three one-parameter families
of two dimensional  Randers-Finlser metrics
discussed in \cite{Crampin}.
There is also some overlap with some of the the spacetimes
described in \cite{Gibbons:2008zi}.
They are  described in detail in
our  notation and conventions in the following
sub-section.

\subsection{The  models} 

The Randers-Finsler metrics described in  \cite{Crampin}
share the feature that they are given in iso-thermal coordinates,
and that in these coordinates the Finsler geodsics
are  circles.

\begin{itemize}

\item $I: \quad \mathbb{S}^2$ 
\ben
a_{ij}dx^ixdx^j = 4\frac{dx^2 + dy^2 }{( 1+ x^2 +y^2 )^2 }
\,,\qquad b_idx^i= 2 \lambda  \frac{ydx-xdy}{(1+ x^2 +y^2 )}    
\,.\een
If
\ben
(x,y)= \tan \frac{\theta}{2} ( \cos \phi ,\sin \phi) \,. 
\een
then
\ben
a_{ij}dx^ixdx^j= d \theta ^2 + \sin ^2 \theta d \phi^2 \,,\qquad b_idx^i= - 2 \lambda\sin^2 \frac{\theta}{2}d \phi \,. 
\een

\item $II: \quad \mathbb{E}^2$ 
\ben
a_{ij}dx^ixdx^j =dx^2 + dy^2 \,,\qquad b_idx^i = \half \lambda (ydx-xdy) \,.   
\een  

If 
\ben
(x,y)=  \rho (\cos \phi, \sin \phi) \,,
\een
then
\ben
a_{ij}dx^idx^j= d \rho ^2 + \rho ^2 d \phi ^2 \,,\qquad b_idx^i= \half \lambda \rho ^2 d \phi \,.
\een

\item $III: \quad  \mathbb{H}^2$
  \ben
  a_{ij}dx^ixdx^j = 4\frac{dx^2 + dy^2 }{( 1- x^2 -y^2)^2  }\,, \qquad
  b_idx^i= 2 \lambda  \frac{ydx-xdy }{(1 -x^2 -y^2 )} \label{hemi} \,.  
\een 
If
\ben
(x,y)= \tanh \frac{\psi}{2} ( \cos \phi ,\sin \phi) \,,
\een
then
\ben
a_{ij}dx^idx^j= d \psi ^2 + \sinh ^2 \psi d \phi^2 \,, \qquad b_idx^i= 2 \lambda \sinh^2 \frac{\psi}{2} d \phi \,.
\een

\end{itemize}

It is case $III$ with $\lambda =1$ which is discussed
in \cite{Markvorsen} 

Note that if $\lambda=1$, then the 2-form $db$ coincides with the
area element on $\mathbb{S}^2$ , $ \mathbb{E}^2$ or   $\mathbb{H}^2$
respectively. Thus the rays coincide with with magnetic geodesics, that
is the trajectories of
charged particles moving in a uniform magnetic field of strength $\lambda$
on $\mathbb{S}^2$ , $ \mathbb{E}^2$ or   $\mathbb{H}^2$
respectively. In particular, for $ \mathbb{E}^2$ we have the Larmor problem,
and  the case of   $\mathbb{S}^2$ is closely connected with the motion of a particle
in the neighbourhood of a magnetic monopole.

Further geometric insight in case $I$   may be afforded by recalling that
if $\tau, \theta, \phi$ are Euler angles for $SU(2)$  then  a left-invariant basis for the its Lie algebra  $\mathfrak{su} (2)$,
is given  by 
\bea
\sigma^1&=& \sin \tau d \theta - \sin \theta \cos \tau d \phi \\
\sigma ^2 &=& \cos \tau d \theta  + \sin \theta \cos \tau d \phi\\
\sigma ^3 &=& d \tau+ \cos \theta d \phi \,.
\eea

The spacetime metric in case $I$ is
\ben
ds ^2_I =-(dt - 2 \lambda \sin^2 \frac{\theta}{2} d \phi )^2  + d \theta ^2 + \sin ^2  \theta d \phi ^2 \,.
\een
let $\tau = \frac{t}{\lambda} -\phi$, then
one finds that
\ben
ds ^2_I  = -\lambda ^2 (\sigma ^3 )^2 + (\sigma ^1)^2 + (\sigma ^2)^2  
\een
which is a left-invariant Lorentzian metric on $SU(2)$ . It's best known appearance in general relativity
is in the Taub-NUT metric. Since the coordinate $\tau$ is necessarily periodic, the spacetime admits closed
timelike curves, CTC's. That is, the spacetime is a model of a time-machine.

Case $II$  may be regarded as a Wigner-Inonu contraction because  
\ben
ds _{II}=   -\bigl (dt + - \half \lambda (ydx-xdy) \bigr )^2  +  dx^2 + dy^2 
\een
is a left-invariant metric on the Heisenberg group,
sometimes call Nill, for which
$dx,dy, dt  - \half \lambda (ydx-xdy) $
provide a left-invariant basis of one-forms.

\subsection{The hemi-spherical elliptic model}

These appear to be  so-called because the fire is envisaged as propagating
on a hemi-spherical shaped piece of   terrain with unit radius
and geometric height function
$z= \sqrt{1-x^2-y^2}$. From the Randers point of view, it is perhaps puzzling
that the metric $a_{ij}$ is not the  metric
induced on the hemi-sphere from the flat metric on Euclidean space $\mathbb{E}^3=(x,y,z)$
which would have constant positive curvature $1$ , but a metric of constant  curvature $-\frac{1}{4}$ 
which in fact is induced on the  top sheet of the  hyperboloid $t= \sqrt{1+x^2 + y^2}$ from the flat
Lorentzian metric of three dimensional Minkowski spacetime $\mathbb{E}^{2,1} - (t,x,y)$.   

The puzzle is resolved when one evaluates the Zermelo metric $h_{ij}$.
This is given by \cite{Markvorsen}
 \ben
h_{ij} dx^i dx^j = \frac{(1-y^2)dx^2 +2 xy dx dy + (1-x^2 ) dy ^2 }{ 1-x^2 - y^2 } \,. 
 \een
 The wind is given
 \ben
W^ i \frac{\p}{\p x^i}  = x \frac{\p}{\p y} - y \frac{\p}{\p x}\,.
 \een

In case $III$ the spacetime metric is 
\ben
ds^2_{III} = -(dt -2\lambda \sinh^2 \frac{\psi}{2} )^2 +
d \psi ^2 + \sinh^2 \psi d \phi ^2 \,, 
\een
If $\lambda = \sqrt{2}$  and one replaces $\frac{\psi}{2} $ by $r$ one
obtains the celebrated  metric of Godel as given in \S $3$  of
\cite{Godel:1949ga}.

More generally if we define  
\bea
\sigma^1&=& \sin \tau d \psi - \sin \psi \cos \tau d \phi \\
\sigma ^2 &=& \cos \tau d \psi  + \sin \psi \cos \tau d \phi\\
\sigma ^3 &=& d \tau+ 2 \sinh^2 \frac{\psi}{2}  d \phi \,,
\eea
we obtain a  left-invariant  basis of one forms  for
$\mathfrak{sl} (2,\mathbb{R})$ 
and the spacetime metric may be expressed as 
\bea
ds ^2_{III}  &=& -\lambda ^2 (\sigma ^3 )^2 + (\sigma ^1)^2 + (\sigma ^2)^2 \\  
&=& -\lambda ^2 (\tau+ \cosh \psi d \phi) ^2 + d \psi ^2 + \sinh^2 \psi d \phi ^2
\label{AdS}\eea
which is a family of  left-invariant Lorentzian metrics on $SL(2,\mathbb{R})$.
If $\lambda^2 =1$, we have, up to a factor of $4$,
the bi-invariant metric on  $SL(2,\mathbb{R})$.
This coincides with three-dimensional anti-de-Sitter spacetime
$AdS_3$ which has the topology $S^2 \times \mathbb{R} ^2 $ and closed
timelike curves. However  if one passes to the
universal covering space $\widetilde{ SL(2,\mathbb{R})  }$
the CTC's are eliminated. By contrast,  if $\lambda^2  >1$
the metrics admit closed timelike curves, even if
one declines to identify the coordinate $\tau$ since $g_{\phi \phi}$
is negative for values of $\psi$. for which $\tanh^2 \frac{\psi}{2} >
\frac{1}{\lambda^2}$. 

To see the concrete relation to more conventional
representations of anti-de-Sitter spacetime define
\ben
Z^1 = \sqrt{1+ r^2}  e^{i T} \,, \qquad Z^2 = r e^{i\Phi}
\een
which embeds $AdS_3 \equiv SL(2,\mathbb{R})$ 
isometrically into
$\mathbb{E}^{2,2}\equiv \{ \mathbb{C}^2 \,,  |d Z^2|^2 - |dZ^1|^2 \} $
as the quadric
\ben
|Z^1|^2  - |Z^2| ^2 =1 \,.
\een
The induced metric is
\ben 
ds^2 = \frac{1}{4} \Bigl \{  -(1+ r^2) dT ^2  +
\frac{dr^2}{1+r^2} + r^2 d \Phi ^2 \Bigr \}  \,. \label{toosmall} 
\een
To recover (\ref{AdS}) multiply the rhs of
(\ref{toosmall}) with $\lambda =1$ by $4$ and set 
\ben
r=\sinh \frac{\psi}{2} \,,\qquad T=\half(\tau-\phi)\,,\qquad \Phi = \half
(\tau+\phi)\,,\quad \phi-\tau =t \,.
\een

To recover (\ref{hemi}) 
define the Lorentzian analogue of  stereo-graphic coordinate $z=x+iy$ by
\ben
z= \frac{Z^2}{Z^1} 
\een
The analogy works as follows. One identifies
the hyperbolic plane with the upper-sheet of the two sheeted
hyperboloid :
\ben
t^2-x^2- y^2 =1 
\een
in three-dimensional Minkowski spacetime  ${\Bbb E}^{2,1} \equiv
\Big \{ \mathbb{R}^3, -dt^2 +dx^2 +dy^2 \Bigr \}$.
One now maps every  point on the upper sheet to the intersection
on the spacelike plane $t=-1$ with the straight line  joining
$(t,x,y)  \,,t\ge 1$     with $( -1,0,0)$.
The reader unfamiliar  with this standard construction
may find more details in the  appendix of \cite{Gibbons:2006mi}.

\subsection{Wave Fronts and Hamilton-Jacobi equation}

Adopting the Zermelo picture,  the Hamilton-Jacobi equation becomes
\ben
h^{ij} \p_ iS \p_j S - (\p_tS + W^i\p_i S) ^2 =0\,,   
\een
which may be rewritten, taking the positive square root,  as
\ben
\p_t S = \sqrt{(h^{ij}\p_iS \p_i S )} -W^i \p_i S  
= G(x,p_i) = \sqrt{h^{ij}p_ip_j} -W^ip_i \label{HJred}
\een
where $p_i=\p_iS$. Now  $G$  coincides with
the expression given in  equation (14) of \cite{Gibbons:2008zi}
for  the moment map  generating  the Zermelo flow on the cotangent space
of the spatial manifold. In fact equation (\ref{HJred})  is in striking
agreement with the Hamilton-Jacobi equation obtained in section \S 2.2
of \cite{SGRMC} by entirely different arguments.

\section{Quantum Navigation}  
There has been interest recently in the quantum control problem of 
finding  a time independent Hamiltonian $\hat H_c$ such that a state 
$|\psi _{\rm initial} \rangle $  of a
system with Hamiltonian $\hat H_0$  is driven to a state
$|\psi _{\rm final} \rangle $  in the shortest possible time subject 
to a bound on ${\rm Tr} {\hat H}_c ^2 $  \cite{R1,R2,Brody:2014voa,BM2,RS3,Brody:2014jaa,Brody:2015zra}.
The Hamiltonian $\hat H_0$ generates the natural
motion or drift  of the system in the absence of external intervention by
an experimenter.  

As we shall show in this section,  the problem is equivalent
to a special case of Zermelo's problem and hence to finding
a Randers-Finsler geodesic. By the theory expounded earlier
this is equivalent to finding a null geodesic in a stationary spacetime. 
In order to understand this connection it is necessary to recast
the standard well known formalism of quantum  mechanics
in which  states correspond to vectors in a Hilbert space
to a more geometrical formulation in which physically distinct
states correspond to points in a particular type of Riemannian manifold
\cite{Strocchi,Cantoni:1977ph,Kibble,Gibbons:1991sa}.
Since  this particular
formalism is less familiar  than it perhaps deserves and we will begin
with a short summary.

\subsection{The space of  physically  distinct quantum states}

It was one of Dirac's greatest achievements to have recognised, shortly after
Heisenberg's discovery  of his eponymous uncertainty relations, the profound
connection between quantum mechanics and Hamiltonian Mechanics.
However not even he realised  that one may regard  quantum mechanics
as a special case of Hamiltonian mechanics \footnote{However 
I owe to Dorje Brody
the observation  that this idea  seems to be presaged in \cite{Dirac}}. 
For that to be possible,
it is necessary to pass from the traditional formulation
of Hamiltonian  mechanics,  which passes by a Legendre transformation
from   Lagrange's equations
\ben
p_a = \frac{\p L}{\p\dot q^a} \,,\qquad \dot p_a =  \frac{\p L}{ \p q^a} 
\een
defined on the $2n$ dimensional  tangent bundle $TQ$
of an $n$-dimensional
configuration
space $Q$ with local coordinates $q^a, v^a= \dot q^a$ 
to  Hamilton's equations  
\ben
\dot q_a = \frac{\p H}{\p p^a} \,,\qquad \dot p_a = -\frac{\p H}{\p q^i} 
\een
defined on the cotangent bundle $T^{\star}Q$   with local coordinates $q^a,p_a$ 
to a more general formulation in which Hamilton's equations are seen
as defining a flow on a $2n$-dimensional  symplectic manifold
$P$ with local coordinates
$x^i $, $i=1,2,\dots 2n $  equipped with
a closed 2-form $\omega_{ij}=-\omega_{ji}$ of maximal rank i.e.
$\det \omega_{ij} \ne 0$, and given by 
\ben
\dot x^i = \omega ^{ij} \frac{\p H}{\p x^i} \,,
\een
where $\omega ^{ij}$  are the components of the inverse of $\omega$. 
We recover the traditional viewpoint if $P=T^{\star}Q$, $x^i=q^q,p_{a+n}$  and 
$\omega = d \theta$, where $\theta = p_adq^a$ is the canonical
one-form.   

Dirac did recognise that the space of physically distinct  quantum states 
is the set of rays, i.e. vectors $|\Psi \rangle \in {\Bbb C} ^{n+1}$
with basis $ \{|n\rangle>, |n+1\rangle \}$,  
in a Hilbert space 
${\cal H}$, which here we take to be of finite dimension $n+1$, 
subject to the equivalence 
\ben
| \Psi \rangle \equiv \lambda | \Psi \rangle  
\een

What Dirac and others of his generation did not seem  to have realised
\footnote{but see previous footnote} is that Sch\"rodinger's equation 
\ben
i \frac{d \,| \Psi \rangle  }{dt} = \hat H  \Psi \rangle 
\een
becomes, when projected onto the space of physically distinct  
quantum states $P$,  
a Hamiltonian flow with respect to a symplectic form 
$\omega$ and Hamiltonian function. 
\ben
H= \langle \Psi |\hat H | \Psi \rangle\, .
\een

In fact $P,\omega$ in this case is complex projective space
${\Bbb C} {\Bbb P}^n \equiv U(n+1)/U(n)\times U(1) $,
which is not only a symplectic manifold, but an Einstein-K\"ahler
manifold. 

To say that  a $2n$-dimensional 
symplectic manifold is K\"ahler is to say
that it admits a Riemannian, i.e. positive definite, metric $a_{ij}$ 
such that if $\nabla_i$   is the associated covariant derivative
operator, then the symplectic form is covariantly constant: 
\ben
\nabla_i \omega _{jk}=0\,,
\een    
and that
\ben
g_{ij} \omega ^{im} \omega ^{jn}= g^{mn} \,.
\een
From this we deduce that $I^i\,_j= g^{ik} \omega _{kj}$  is covariantly
constant and  satisfies 
\ben
I^i\,_kI^k\,_j = - \delta ^i_j \,.
\een
It follows that $I^i\,_j$ is an integrable complex complex structure
for the manifold $P$, in other words we may consistently introduce charts
of $n$ complex coordinates with overlap functions which are
locally holomorphic to cover our manifold. To express things more briefly
a K\"ahler manifold is one whose holonomy group is contained within
$U(n) \subset SO(2n)$.  
Finally, to say that a Riemannian
manifold  is Einstein is to say that the the Ricci tensor
satisfies
\ben
R_{ij}= \Lambda g_{ij}\,,
\een
for some constant $\Lambda$.

The standard  flat K\"ahler structure on ${\Bbb C} ^{n+1}$
is given by
\ben
ds ^2 = | dZ ^k|^2  \,,\qquad  \omega = 
\frac{1}{i}  d Z^k \wedge d {\bar Z} ^k = dp _k \wedge d q^k 
\een
where $Z^k= \frac{1}{\sqrt2}(q^k  + i p^k) $.
The space of unormalised states is thus a flat K\"ahler manifold
with 
\ben
ds^2 = \langle d\Psi |d  \Psi \rangle  \,.
\een
where
\ben
|\Psi \rangle = Z^k |k \rangle 
\een
and $\{|k \rangle \}$ is an orthonormal basis.

One may check   that Schr\"odinger's equation may be written
as 
\ben
i\frac{d Z^k}{dt} = \frac{ \p H (Z,\bar Z)  }{\p {\bar Z}^k} 
\een
or 
\ben
\frac{d q^k}{dt}= \frac{\p H (p,q) }{\p p_k} \,,\qquad \frac{d p_k }{dt}= 
- \frac{\p H(p,q) }{\p q^k} 
\een

The general K\"ahler manifold admits local complex coordinates $Z^k$  
and a so-called K\"ahler potential $K ( Z, {\bar Z} ) $ such that 
\ben
ds ^2 = \frac{\p ^2 K}{ \p Z^m \p {\bar Z}^n } 
 d  Z ^m \otimes _s  d {\bar Z}  ^n \,,
\qquad \omega =  \frac{1}{i} \,  \frac{\p ^2 K}{ \p Z^m \p {\bar Z} ^n }
\,  dZ^ m \wedge d { \bar Z} ^n \,,  
\een  
For the flat K\"ahler structure on ${\Bbb C} ^{n+1}$ we have
\ben
K= Z^k {\bar Z} ^k  = <\Psi | \Psi > \,.
\een

Because a K\"ahler manifold is a symplectic  manifold,
it also admits local Darboux coordinates such that
\ben
\omega = dp_k \wedge  d q^k \,.
\een  

In the case of the flat coordinates 
used above, they are both complex and Darboux.
For a general K\"ahler manifold this need  not be the case.  

We now give a construction of the Fubini-Study metric
on ${\Bbb C}{\Bbb P}^n$ the space of physically distinct
quantum states.   We first 
normalise our state vectors
\ben
\langle \Psi | \Psi \rangle =1.\label{norm}
\een
The space of normalised  states may be identified
with the $2n+1$ sphere $S^{2n+1} = U(n+1)/U(n)$,
 with its standard unit round metric 
\ben
ds^2_{2n+1} = \langle d\Psi | d\Psi \rangle\,. \label{round}
\een    
Moreover $U(n) \subset SO(2n) $ acts transitively and isometrically.
We  now eliminate the remaining phase freedom in (\ref{norm}).
Thus we restrict $\lambda$ to be an element of $U(1)$,
\ben
\lambda = e^{i\psi} \,,
\een
and identify points on $S^{2n+1}$ under the action of this $U(1)$.
Thus ${\Bbb C}{\Bbb P}_n \equiv S^{2n+1}/U(1) $. This construction
of $S^{2n+1}$ as a $U(1)$ bundle over $ {\Bbb C}{\Bbb P}^n$
is known to topologists as the Hopf fibration. The obits of $U(1)$
in $S^{2n+1}$ are called the Hopf fibres.

To obtain the $SU(n)/{\Bbb Z}_n$ invariant  
Fubini-Study metric $g_{ab}(x)$  on ${\Bbb C}{\Bbb P}^n$
we project the round metric (\ref{round}) orthogonally
to the Hopf fibres. Thus  
\ben
ds ^2_{2n+1}  = \frac{1}{4} 
(d \psi + A_a dx^a) ^2 + g_{ij}(x^c) dx^i dx ^j
\een
where $i=1,2,\dots 2n$ and find that ({\bf up to factor?}) that
locally the K\"ahler form is given by
\ben
\omega = d A_i dx^i \,,  
\een
and if  
\ben
g_{ij}dx^i dx^j = \langle d\Psi |d\Psi \rangle -
|\langle \Psi | d \Psi > |^2\,.\een 

For more details of this general formalism and how
it relates to the language of q-bits 
and  their entanglement the reader may consult
\cite{Cvetic:2015uga}. 

\subsubsection{The Bloch Sphere}

If $n=2$ the   space of physically distinguishable states is
known as the Bloch sphere \cite{Bloch}.
Geometrically this is the same as a sphere of radius $\half$.
Thus If $Z^1,Z^2$ are affine coordinates for ${\Bbb C}^2$
then  $S^2 \equiv  {\Bbb C \Bbb P}^1  = \{ (Z^1,Z^2)  \,| (Z^1,Z^2
\equiv  \lambda (Z^1,Z^2), \lambda \in {\Bbb C} ^\star \} $, 
and recalling  that $S^3 \equiv SU(2)$  
we may parameterize then in terms of a radius $r=\sqrt{|Z^1|^2 + |Z^2|}$ 
and Euler angles $\{ (\psi, \theta ,  
\phi)  \, |  0\le \psi < 4 \pi, 0\le \theta < \pi,
0 \le \phi < 2 \pi \} $   as 
\bea
Z^1 &=& r e^{\frac{i}{2} (\psi + \phi)} \cos\frac{\theta}{2} \\
Z^2 &=& r e^{\frac{i}{2} (\psi - \phi)} \sin  \frac{\theta}{2} 
\eea
The Hopf fibration  $S^3 \rightarrow  {\Bbb C  \Bbb P}^1$
is given by  $\{ (\psi, \theta ,  
\phi) \rightarrow (\theta, \phi) $   
and the Hopf fibres 
are given by $r=1$, $ \theta, \phi ={\rm constant} $
The metric on ${\Bbb C} ^2 \equiv {\Bbb E} ^4 $ is
\ben
ds ^2 = |dZ ^1|^2  + |d Z ^2|^2  = dr ^2 + \frac{r^2}{4} 
\Bigl( (d \psi + \cos \theta d \phi ) ^2 +  
d \theta ^2 + \sin ^2 \theta  d \phi ^2 \Bigr ) 
\een 
The inhomogeneous coordinate on $ {\Bbb C  \Bbb P}^a $ is 
\ben
\zeta = \frac{Z^1}{Z^2} =  e^{i \phi}   \cot \frac{\theta}{2}
\een
which geographers call stereographic coordinates
in terms of which the Fubini-Study metric on $ {\Bbb C  \Bbb P}^1 $ is
\ben
ds ^2 = \frac{ |d \zeta|^2 }{ ( 1+ |\zeta |^2 ) ^2 } =  
\frac{1}{4} (d \theta ^2  + \sin ^2  \theta d \phi  ^2) 
\een
Evidently stereographic coordinates do not provide a Darboux chart
since the the area element  is
\ben
 \frac{i}{2} \frac {d \zeta \wedge d \bar \zeta} {(1+|\zeta|^2 ) ^2 } \,.  
\een  

However if we  define
\ben
a= \frac{\zeta}{\sqrt{1+ |\zeta|^2 }} \,, \qquad \Longleftrightarrow \qquad
\zeta =  \frac{a}{\sqrt{1- |a|^2 }} \,,
\een  
we find 
\ben
 \frac{i}{2} \frac {d \zeta \wedge d \bar \zeta} {(1+|\zeta|^2 ) ^2 } 
=  \frac{i}{2} da \wedge d \bar a \,. \label{trans}
\een
Thus $a= p+iq = e^{i\phi} \cos(\frac{\theta}{2})  
$ is a Darboux chart for ${\Bbb C}{\Bbb P} ^1$.
and the total area of ${\Bbb C}{\Bbb P} ^1$  with respect to its
Fubini-Study metric is 
\ben
\int _{|a|^2 \le 1} dp dq = \pi =  \frac{1}{4} 4 \pi \,,
\een as expected. 
Geographers call   the  map from $S^2$ to the unit disc
 {\it Lambert's Polar Azimuthal
Equal Area Projection}
Note that while this projection  is a symplecto-morphism
or canonical transformation,
i.e. preserves the symplectic form, and hence the area element,  but it does not
preserve the complex structure, in other words $a$ is not a locally  
holomorphic function of $\zeta$.  
 
\subsubsection{  ${\Bbb C}{\Bbb P} ^n$ }

The previous theory generalises in an almost trivial way to an $n+1$ state
system.
One introduces $n$ inhomogeneous coordinates  $\zeta^a\,, a=1,2,\dots n $ 
for $ {\Bbb C}{\Bbb P} ^n$. The K\"ahler form is 
\ben
\omega =\frac{i}{2}\, \p_{\zeta ^a }\p_{{\bar \zeta }^a} K  d \zeta ^a \wedge d {\bar \zeta }^a  \,,\qquad K= \ln(1+ 
{\bar \zeta} ^a  \zeta^a )\,, 
\een
 and Darboux coordinates are 
\ben
a^a= \frac{\zeta^a}{\sqrt{1+ |\zeta|^2 }} \,, \qquad \Longleftrightarrow \qquad
\zeta^a =  \frac{a^a}{\sqrt{1- |a|^2 }} \,,
\een  
with $|z|^2 = z^a {\bar z}^a$ etc and  the Einstein summation convention is 
adopted.   If  $a^a=p_a + \sqrt{-1}q^{n+a}j$ we have, in this Darboux chart  
\ben
\omega = dp_a \wedge dq ^a\,.
\een

Thus a random quantum state, i.e the perfectly ignorant density matrix,
consists  of quantum states  uniformly distributed  inside the unit ball
$  (p_a) ^2 + (q^a) ^2 \le 1 $  and Schr\"dinger's equation  is quite
literally Hamilton's original  equation in thse coordinates. 
It is of interest 
to write out the general Hamiltonian function in terms of $p_a$ and $q^a$
,not least, since  the system is clearly integrable : the states evolve under a
one parameter subgroup of $SU(n+1)$.

In more detail we have
\ben
|\Psi \rangle = \frac{1}{\sqrt{1+|\zeta|^2}} \Bigl (\zeta ^a |a \rangle 
+ |n+1\rangle \Bigr ) =  a ^a |a \rangle 
+  \sqrt{1- |a|^2}               |n+1\rangle 
\een

\ben
H= \langle \Psi |\hat H |\Psi \rangle
=  {\bar a} ^a H_{ab} a^b +  (1- |a|^2 ) H _{(n+1) \,(n+1) }
+ \sqrt{1-|a|^2 } \bigl( {\bar a}^a H_{a\,(n+1)}  + H_{(n+1)\,a} a^a \bigr )         \een

\subsection{Extended Phase Space}

In both  classical mechanics and quantum mechanics the 
time coordinate $t$ and the phase  space 
coordinates $x^a$ are  on a very different
footing. Firstly one is allowed to use
arbitrary coordinates on the symplectic manifold 
$\{P,\omega\} $, j
Just as one is allowed to use arbitrary coordinates on a 
Riemannian or pseudo-Riemannian manifold $\{M,g\}$.
Moreover by analogy to isometries, which are diffeomorphisms
 $f:M\rightarrow  M$  under which the metric is invariant
\ben
f_\star g=g \,,
\een
where $f_\star$ denotes the pull-back map, one defines
symplectomorphisms or     canonical transformations
as  coordinate transformations $f : P\rightarrow P$ 
which leave invariant the symplectic form $\omega$,
\ben
f_\star \omega =\omega    \,.
\een
By contrast the time coordinate is regarded as ``absolute''
and one typically does not consider coordinate transformations 
mixing $t$ and $x^a$. 

This gives rise to no significant  difficulties  if the Hamiltonian $H$
is independent of time and Hamilton's equations
are an autonomous set of first order ordinary differential equations
on $P$  but it becomes inconvenient if one 
is considering, as one does in control theory, time dependent Hamiltonians,
when Hamiltonians are non-autonomous, or when one wishes to make use
of symmetries of Hamilton's  equations, 
such as Galilei or Lorentz transformations which mix $x^a$ and $t$.

To this end, on may pass to a $2n+1$ dimensional 
extended phase space, sometime called
Evolution Space \cite{Souriau},  $V= P \times {\Bbb R} $ with coordinates
$X^\alpha  = t, x^a$   , equipped with
a closed so-called pre-symplectic 2-form   
\ben
\Omega  = \omega -dH \wedge dt \,.
\een
The 2-form   $\Omega$  is pre-symplectic rather than
because  it has a kernel  $V^\mu$,  
\ben
\Omega _{\alpha \beta } V^\beta   = 0 
\een 
where $V^\beta $ is the tangent vector to the 
lift to $E$ of the solutions $t=t(\lambda)$, $ x^a = x^a(\lambda)$ 
 of Hamilton's equations, that is 
\ben
V^0 = \frac {dt}{d \lambda} \,,\qquad V^a = \frac{d x^a}{d \lambda}\,.
\een

In the special case that the symplectic manifold
is a co-tangent bundle of some configuration space 
$P= T^\star Q$, then
\ben
\Omega = d \Theta \,,\quad \Theta = \theta -H dt = p_idq^i - H dt
\een 
and   $\{E,\Theta\} $ is a a special case of a  contact manifold
with globally defined contact 1-form $\Theta$, which by definition
has exterior derivative $d \Theta$  of maximum rank.
Diffeomorphisms $f: E \rightarrow  E$ such that
\ben
f_\star \Theta = \Theta 
\een
 are called   ``contact transformations'' or ``contacto-morphisms''.

Thus if $Q={\Bbb E} ^3$ and $H=\frac{1}{2m} \bp ^2 $ 
or $H=\sqrt{\bp ^2 c^2 + m^2 c^4}$, one may check that  
Galilei or Lorentz transformations are contacto-morphisms respectively.
 
\subsection{Extended Quantum Statespace as a Stationary Spacetime}

For our present purposes we may  choose
$h_{ij}$ to be the Fubini-Study metric on  ${\Bbb C}{\Bbb P}^n$ 
and $W^i$ the vector field generating the drift.
In other words, $W^i$ is a Killing wind
generating the one parameter subgroup of $SU(n+1)$
with moment map
\ben
\langle\Psi  |\hat H_0 | \Psi \rangle \,.  
\een

Thus we see that \emph{the time-independent quantum control
problem is solved by null geodesics moving of the 
stationary spacetime  on the extended quantum phase space 
$ {\Bbb C}{\Bbb P}^n \times {\Bbb R} $ equipped with the 
 Lorentzian  metric (\ref{Lorentz}).}

\subsubsection{Example: a  spin  $\half$ system on the Bloch Sphere }

The spin is caused to precess with the Larmor frequency $\omega_L$
by an external magnetic field. Thus   
\ben
h_{ij}dx^i dx^j= 
\frac{1}{4} \Bigl \{ d \theta ^2 + \sin ^2 \theta d \phi ^2 \Bigr \} 
\,,\qquad W^i = \omega _L \delta ^i _\phi \,.
\een
and the stationary metric (\ref{Zspacetime}) becomes 
\ben
ds ^2 =-dt ^2 +  \frac{1}{4} \Bigl \{ d \theta ^2 + 
\sin ^2 \theta ( d \phi- \omega_L dt )  ^2 \Bigr \}\,.
\een
Since $\phi$ is ignorable, we set $F=h \phi + \Theta(\theta)$ and 
find 

\ben
\bigl(\frac{d \Theta }{d \theta }\bigr ) ^2 + \frac{h^2}{\sin ^2 \theta}
= 4 ( \cE - h \omega_L) ^2 \,.    
\een
 
We then have
\bea
\frac{d \phi}{d \lambda} &=& \frac{4 h}{\sin ^2  \theta}\\  
\frac{d \theta} {d \lambda } &=& \pm \sqrt{ 4 ( \cE - h \omega_L) ^2 - \frac{h^2}{\sin ^2 \theta}   }   \\
\frac{dt}{d \lambda} &=&  ( \cE - h\omega_L ) \,. 
\eea

It is clear that the problem is equivalent to the usual problem but with the
replacement
\ben
\phi \rightarrow \tilde \phi = \phi-\omega_L t \,,  
\een 
 which corresponds to the device of passing  to the interaction picture.
adopted in \cite{Brody:2014jaa} .

\subsubsection{Example a spin  $1$ system on ${\Bbb C}{\Bbb P} ^2$  }

${\Bbb C}{\Bbb P} ^2$ first came to the attention of physicists
as a ``gravitational instanton'' \cite{Eguchi:1976db}.
Later it was recognised as the state space of a spin one system
\cite{Bouchiat:1987vf}

If one uses a coordinate system adapted to the action of $U(2)$ subgroup
of the $SU(3)/{\Bbb Z}_3$ subgroup of the isometry group
of ${\Bbb C}{\Bbb P} ^2$, the Fubini-Study metric is \cite{Gibbons:1978zy}
\ben
h_{ij}dx^i dx^j = \frac{dr^2}{(1+r^2 )^2}  + \frac{r^2}{4 (1+r^2 )^2}
(d \psi + \cos \theta d \phi ) ^2 +  \frac{r^2}{4 (1+r^2 )} \bigl(  d \theta ^2 + \sin ^2 \theta d \phi ^2 \bigr ) \,.
\een
The coordinates $\psi,\theta,\phi$ are Euler angles  on $SU(2)$ 
and $0\le r \le \infty$. Introducing $u=1/r$ we can add a 
${\Bbb C}{\Bbb P} ^1$   at infinity
to obtain a two-chart atlas covering the entire, compact manifold. 
Alternatively,   if $r=\tan \chi$, $0\le \chi \le \frac{\pi}{2}$, we have  
\ben
h_{ab}dx^a dx^b = d \chi ^2   + \frac{1}{4} \Bigl \{ 
\sin ^2 \chi \cos ^2 \chi   (d \psi + \cos \theta d \phi ) ^2 +  \sin^2 \chi \bigl(  d \theta ^2 + \sin ^2 \theta d \phi ^2 \bigr ) \Bigr \} \,.
\een

If $|\Psi\rangle = Z^I |\Psi _I \rangle \,,$ 
where $|\Psi _I \rangle $ are an orthonormal basis of states,
$ \langle \Psi _i|  \Psi _j \rangle = \delta_{IJ}$,  
and if $Z^3 \ne 0$, we may parametrise the space of distinguishable 
quantum states
as    
\bea
\frac{Z^1}{Z^3}&=&  \tan \chi \cos (\frac{\theta}{2}) e^{\frac{i}{2}(\psi + \phi)}\\ 
\frac{Z^2}{Z^3}&=&  \tan \chi\sin (\frac{\theta}{2}) e^{\frac{i}{2}(\psi - \phi)}\,.
\eea

There are two commuting Killing vectors $\p_\psi$ and $\p_\phi$ 
which suggest  that we choose for the wind
\ben
W^i \p_i = \omega _\psi p_\psi + \omega _\phi \p_\phi \,.
\een    

Note that $SU(3)$ is of rank $2$ and therefore  this 
may  be the most general choice.

We assume that 
\ben
S= -\cE t +h_\psi \psi +  h_\phi  \phi + W \,.
\een
and  find that  
\ben
 4 (\frac{\p W}{\p \chi })^2 +  
\frac{4} {\sin^2 \chi } \Bigl \{(
\frac{\p W}{\p \theta })^2
+ \frac{1}{\sin^2 \theta}
 (h_\phi - \cos \theta h_\psi) ^2 \Bigr \} 
+  \frac{4}{ \sin ^2 \chi \cos ^2 \chi}    h_\psi^2 = \bigl( \cE - \omega _\psi h_\psi -\omega _\phi h_\phi \bigr)^2         \,.  
\een

This clearly separates. We set $W= X(\chi) + \Theta (\theta)$ and 
require 
\ben
( \frac{d\Theta }{d \theta })^2
+ \frac{1}{\sin^2 \theta}
 (h_\phi - \cos \theta h_\psi) ^2 = h^2 
\een
and 
\ben (\frac {d X}{d \chi } )^2 + 
\frac{4h^2 } {\sin^2 \chi }  
+  \frac{4 h_\psi ^2 }{ \sin ^2 \chi \cos ^2 \chi} = \bigl( \cE - \omega _\psi h_\psi -\omega _\phi h_\phi \bigr)^2   \,.   
\een

\section{Acknowledgements} 

I would like to thank Dorje Brody, Christian Duval, Peter Horvathy, Steen Markvorsen,  David  Meier and Claude Warnick  for helpful
discussions and comments on various aspects of the  the material of this paper.

\end{document}